\documentclass[lettersize,journal]{IEEEtran}
\usepackage{amssymb}
\usepackage{array, makecell}
\usepackage{amsmath,amsfonts}
\usepackage{algorithmic}
\usepackage{algorithm}
\usepackage{mathtools}
\usepackage{array}
\usepackage[caption=false,font=normalsize,labelfont=sf,textfont=sf]{subfig}
\usepackage{hyperref}
\usepackage{textcomp}
\usepackage{xcolor}
\usepackage{stfloats}
\usepackage{url}
\usepackage{verbatim}
\usepackage{graphicx}

\usepackage{cite}

\newcommand\myeqa{\stackrel{\mathclap{\normalfont\mbox{(a)}}}{=}}

\newcommand\undermat[2]{%
  \makebox[0pt][l]{$\smash{\underbrace{\phantom{%
    \begin{matrix}#2\end{matrix}}}_{\text{$#1$}}}$}#2}

\begin{document}
\title{A Neural Network-Prepended GLRT Framework for Signal Detection Under Nonlinear Distortions}


\author{Rajeev~Sahay,~\IEEEmembership{Graduate Student Member,~IEEE,}
        Swaroop~Appadwedula,~\IEEEmembership{Member,~IEEE,}
        David~J.~Love,~\IEEEmembership{Fellow,~IEEE,}
        and~Christopher~G.~Brinton,~\IEEEmembership{Senior~Member,~IEEE}
\thanks{R. Sahay, D. J. Love, and C.G. Brinton are with the Elmore Family School of Electrical and Computer Engineering, Purdue University, West Lafayette, IN, 47907 USA. E-mail: \{sahayr,djlove,cgb\}@purdue.edu.}
\thanks{S. Appadwedula is with MIT Lincoln Laboratory, Advanced SATCOM Group, Lexington, MA, 02421 USA. E-mail: swaroop@ll.mit.edu.}
\thanks{This work was supported in part by MIT Lincoln Laboratory.}
\thanks{DISTRIBUTION STATEMENT A. Approved for public release. Distribution is unlimited. This material is based upon work supported by the Department of the Navy under Air Force Contract No. FA8702-15-D-0001. Any opinions, findings, conclusions or recommendations expressed in this material are those of the author(s) and do not necessarily reflect the views of the Department of the Navy.}
}



\maketitle

\begin{abstract}
Many communications and sensing applications hinge on the detection of a signal in a noisy, interference-heavy environment. Signal processing theory yields techniques such as the generalized likelihood ratio test (GLRT) to perform detection when the received samples correspond to a linear observation model. Numerous practical applications exist, however, where the received signal has passed through a nonlinearity, causing significant performance degradation of the GLRT. In this work, we propose prepending the GLRT detector with a neural network classifier capable of identifying the particular nonlinear time samples in a received signal. We show that pre-processing received nonlinear signals using our trained classifier to eliminate excessively nonlinear samples (i) improves the detection performance of the GLRT on nonlinear signals and (ii) retains the theoretical guarantees provided by the GLRT on linear observation models for accurate signal detection.

\end{abstract}


\section{Introduction}


Signal detection is a classical problem with several applications in wireless communications \cite{detect1}. The general detection problem entails identifying the presence of a particular signal of interest (SOI) in a received transmission containing different sources of background noise and interference. Traditionally, with adequate assumptions about the distributions of the SOI and interference, statistical hypothesis testing has been used to derive robust and highly accurate detectors \cite{klrt,aglrt}. 

Despite their robustness under classical linear operations, nonlinear transformations and clipping of a received signal significantly degrades the performance of linear test statistics derived from statistical hypothesis testing. In particular, and of special interest to our study, high-powered throughput signals that are outside the linear operating regime of an amplifier cause nonlinear signal distortions, which result in clipped signals leading to nonlinearities in the received passband signal before it is observed at complex baseband. Such high-powered signals that induce nonlinearity arise from sources of interference 
such as loud signals in audio applications and high signal-to-noise ratio (SNR) interferers that dominate the SOI in situations such as the near-far problem. Achieving accurate detection in such scenarios despite the nonlinearity caused by the system's hardware is vital. Therefore, robust detection models, with performance guarantees similar to those provided by likelihood methods, are needed for signal detection in the presence of hardware-induced nonlinearities.

In this work, we consider editing a received wireless signal that has undergone a nonlinear transformation prior to applying the generalized likelihood ratio test (GLRT) to perform detection. We find that, under nonlinear distortions of a received signal, each sample has been altered to a different extent. For example, certain nonlinear samples are significantly distorted from their linear counterparts while other nonlinear samples remain close to their linear counterparts. Motivated by this, we train a lightweight (i.e., low-parameter) feedforward dense neural network (DNN) capable of classifying samples that have undergone excessive nonlinearity. At test time, we pre-process our received nonlinear signal by eliminating the time samples that are predicted to be highly distorted by the DNN. The GLRT is then applied to the pre-processed signal, with the highly nonlinear time samples eliminated, to perform detection. Our results verify that this method allows a linear test statistic to operate in nonlinear environments while also retaining the theoretical guarantees of the GLRT.

\textbf{Related Work:} Historically, test statistics derived from the GLRT, as well as similar statistical hypothesis testing methods \cite{ofdm_nl_clip2}, have been thoroughly studied and shown to be robust signal detectors in linear operations \cite{glrt1,glrt2} with Gaussian noise. In addition, the GLRT has also been shown to be effective when the distribution of the background noise in the received signals is non-Gaussian \cite{nl_metric1,nl_metric2}. Yet, the GLRT is prone to significant performance degradation when the received signal has undergone a nonlinear transformation, regardless of the noise distribution \cite{nl_dist}. 


Contrary to the long-standing work on detection using likelihood methods, machine learning has recently been proposed for signal detection in wireless communications and cognitive radio environments \cite{nl_clipping,ofdm_nl_clip,adv_amc}. In particular, these studies have demonstrated the versatility that deep learning is able to provide in dynamic environments over linear methods when analytically characterizing the received signal may not be possible. However, contrary to likelihood-based detection, deep learning does not provide theoretical guarantees for detection, making the interpretation of the detection result challenging. In this work, we employ neural networks in conjunction with a likelihood-based detection test to leverage its strong classification performance while retaining the theoretical guarantees provided by the GLRT. 


\textbf{Summary of Contributions:} The main contributions of this work are as follows: 

\begin{itemize}
    
    \item \textbf{Nonlinear Sample Classifier}: We propose prepending a GLRT detector with a neural network classifier to identify time samples that have undergone a nonlinear transformation (Sec. II).
    
    \item \textbf{Linear GLRT Operation in Nonlinear Environment}: We show the effectiveness of our nonlinear sample classifier, in conjunction with the GLRT-derived linear test statistic, in performing robust detection in nonlinear environments (Sec. III). 
    
        
    
    
\end{itemize}

\section{Methodology}



\subsection{Signal Modeling}

\begin{figure*}[htb] 
	\centering
	\includegraphics[width=2.0\columnwidth]{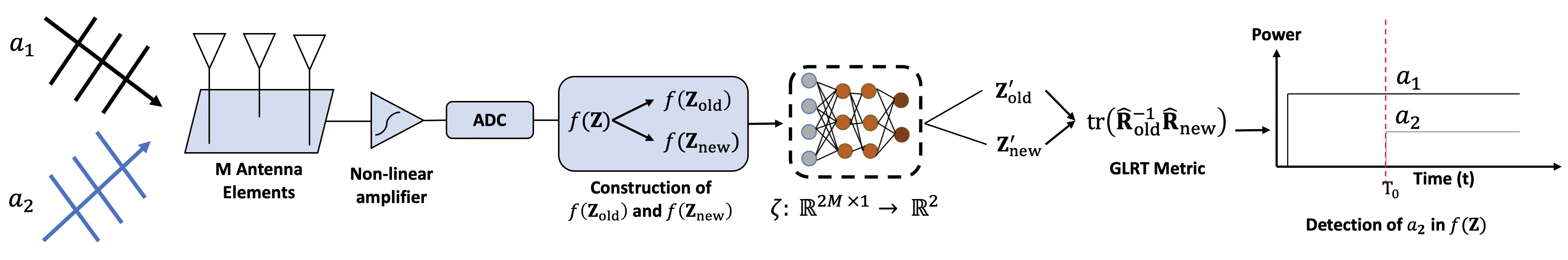}
	\caption{Our detection framework for nonlinear signals consisting of a DNN prepending the GLRT. Our framework is designed for the detection of a SOI, $a_{2}$, when the received signals have undergone nonlinearity due to hardware constraints.}
	\label{block_diagram}
\end{figure*}

We consider two time-series signals, $a_{1}$ and $a_{2}$, transmitted from two different directions, impinging an $M$ element antenna array. Throughout this work, \emph{we will denote the interference signal as $a_{1}$ and the SOI, which we are interested in detecting, as $a_{2}$}. A snapshot of the output at the antenna array at time $t$ is given by
\begin{align}
    \mathbf{z}(t) = \mathbf{x}(u_{1})a_{1}(t) + \mathbf{x}(u_{2})a_{2}(t) + \mathbf{n}(t), \quad  &t = 1, \ldots, L , 
\end{align} 
where $\mathbf{z}(t) \in \mathbb{C}^{M \times 1}$ is a column vector that denotes the time sample of the $M$ sensor outputs at time $t$, $\mathbf{x}(u_{1})$ and $\mathbf{x}(u_{2})$ are the array response vectors of signals $a_{1}$ and $a_{2}$ arriving from directions $u_{1}$ and $u_{2}$ (assuming that the direction parameters $u_{1}$ and $u_{2}$ are scalars), respectively, and $a_{1}(t)$ and $a_{2}(t)$ are the amplitudes of $a_{1}$ and $a_{2}$, respectively, at time $t$. Lastly, $\mathbf{n}(t) \sim \mathcal{C}\mathcal{N}(\mathbf{0}, \sigma^{2}\mathbf{I})$ is complex additive white Gaussian noise (AWGN) with variance $\sigma^{2}$, and $L$ is total number of time samples.

Let $\mathbf{Z} = [\mathbf{z}(1), \cdots, \mathbf{z}(L)] \in \mathbb{C}^{M \times L}$ denote the complete observed signal at the antenna array. Given $\mathbf{Z}$, we are interested in determining when the SOI turns on (i.e., when $a_{2}$ becomes active). Here, we define the null hypothesis, $\mathcal{H}_{0}$, as the case where $a_{2}$ is absent in $\mathbf{Z}$ (i.e., $a_{2} = \mathbf{0}$), and the alternate hypothesis, $\mathcal{H}_{1}$, as the case where $a_{2}$ is present in $\mathbf{Z}$ (i.e., $a_{2} \neq \mathbf{0}$). Formally, we characterize the distribution of $\mathbf{Z}$ under each hypothesis as
\begin{equation} \label{hyp_def}
  \mathbf{Z} \sim
    \begin{cases}
      \mathcal{C}\mathcal{N}(\mathbf{0}, \mathbf{R}), & \mathcal{H}_{0}\\
      \mathcal{C}\mathcal{N}(\mathbf{XAT}, \mathbf{R}), & \mathcal{H}_{1} 
    \end{cases},        
\end{equation}
where $\mathbf{R}$ is the true, unknown covariance of $\mathbf{Z}$. Now, under $\mathcal{H}_{1}$, we are exclusively considering a system with one SOI in $\mathbf{Z}$. Here, $\mathbf{X} = [\mathbf{x}(u_{2})] \in \mathbb{C}^{M \times 1}$ is the antenna array response of the SOI, $\mathbf{A} \in \mathbb{C}^{1 \times S}$ is the row vector of signal amplitudes, and $\mathbf{T} \in \mathbb{C}^{S \times L}$ is the matrix $[\mathbf{0} \quad \mathbf{I}_{S}]$ ($S = L - T_{0}$) corresponding to the time samples during which the SOI is active (i.e., when $a_2$ is present in $\mathbf{Z}$), thus yielding $\mathbf{AT} = [0, 0, \ldots, 0, a_{2}(T_{0}), a_{2}(T_{0}+1), \ldots, a_{2}(L)]$, where $T_{0}$ denotes the time sample at which $a_{2}$ becomes active in $\mathbf{Z}$. Our objective is to arrive at a metric that will allow us to determine the active hypothesis (i.e., where the SOI became active in $\mathbf{Z}$).

\subsection{GLRT-based Linear Detection}

Given $\mathbf{Z}$, we will determine the active hypothesis by utilizing the GLRT and optimizing over the empirical covariances. Denoting $h_{0}$ and $h_{1}$ as the complex Gaussian densities corresponding to the distributions under $\mathcal{H}_{0}$ and $\mathcal{H}_{1}$, the GLRT, by definition, is 
\begin{equation}
    \frac{\underset {\mathbf{R}, \mathbf{A}}{\text{max}} \hspace{1mm} h_{1}(\mathbf{Z}; \mathbf{R}, \mathbf{A})} {\underset {\mathbf{R}}{\text{max}} \hspace{1mm} h_{0}(\mathbf{Z}; \mathbf{R})} \myeqa \frac{|\hat{\mathbf{R}}_{0}|} {\underset {\mathbf{A}}{\text{min}} \hspace{1mm} |\hat{\mathbf{R}}_{1}|} \nonumber
\end{equation}
\begin{equation} \label{glrt_int}
    = \frac {|\mathbf{ZZ}^{\text{H}}|} {\underset {\mathbf{A}}{\text{min}} \hspace{1mm} |(\mathbf{Z} - \mathbf{XAT}) (\mathbf{Z} - \mathbf{XAT})^{\text{H}}|} \geq \gamma, 
\end{equation}
\noindent where $\gamma$ is the threshold parameter, $\text{H}$ denotes the Hermitian transpose, $\text{tr}(\cdot)$ denotes the trace of a matrix, and $|\cdot|$ denotes the determinant. In (a), the maximization over the numerator and denominator covariances yield the corresponding empirical covariances $\hat{\mathbf{R}}_{0} = \mathbf{ZZ}^{\text{H}}$ and $\hat{\mathbf{R}}_{1} = (\mathbf{Z} - \mathbf{XAT}) (\mathbf{Z} - \mathbf{XAT})^{\text{H}}$. The quantity in (\ref{glrt_int}) is equivalent to
\begin{equation} \label{glrt2} 
    \frac {|\mathbf{X}^{\text{H}} (\mathbf{Z}_{\text{old}} \mathbf{Z}_{\text{old}}^{\text{H}})^{-1} \mathbf{X}|} {|\mathbf{X}^{\text{H}} (\mathbf{Z} \mathbf{Z}^{\text{H}})^{-1} \mathbf{X}|} \geq \gamma.
\end{equation}
We refer the reader to \cite{glrt_deriv} for additional details on this GLRT derivation. In (\ref{glrt2}), we have introduced $\mathbf{Z}_{\text{old}} \in \mathbb{C}^{M \times k}$, which corresponds to a partition of subsequent time samples in $\mathbf{Z}$. This definition assumes that the GLRT will be applied to partitions of $\mathbf{Z}$. Specifically, $\mathbf{Z}_{\text{old}}$ will be used in conjunction with the next partition of time samples, $\mathbf{Z}_{\text{new}} \in \mathbb{C}^{M \times k}$, to determine if a new signal (the SOI) became active in $\mathbf{Z}_{\text{new}}$ which was absent in $\mathbf{Z}_{\text{old}}$ \cite{glrt2}. Formally, $\mathbf{Z}_{\text{old}}$ and $\mathbf{Z}_{\text{new}}$ are given by
\begin{equation} \nonumber
\mathbf{Z} = \begin{bmatrix}
\undermat{\mathbf{Z}_{\text{old}}}{\mathbf{z}(1), \cdots, \mathbf{z}(k),} \undermat{\mathbf{Z}_{\text{new}}}{\mathbf{z}(k+1), \cdots, \mathbf{z}(2k),} \cdots, \mathbf{z}(L)\\
\end{bmatrix}, 
\end{equation} 
\vspace{0.25cm}

\noindent where $\mathbf{Z}_{\text{old}}$ and $\mathbf{Z}_{\text{new}}$ are shifted one time sample per iteration, and the GLRT is reevaluated, for all time samples. 

In our case, where $\mathbf{X}$ is completely unknown, we can base detection on the equivalent monotonically related suboptimal statistic of the GLRT \cite{glrt2} given by 
\begin{equation} \label{detection}
    \text{tr}({\hat{\mathbf{R}}_\text{old}}^{-1}\hat{\mathbf{R}}_{\text{new}}) \geq \gamma, 
\end{equation}
where we will reject $\mathcal{H}_{0}$ if $\gamma \geq \gamma_{0}$ for some threshold $\gamma_{0}$, which is empirically determined. Here, $\hat{\mathbf{R}}_{\text{old}} = \mathbf{Z}_{\text{old}}\mathbf{Z}_{\text{old}}^{\text{H}}$ is the empirical covariance of $\mathbf{Z}_{\text{old}}$, and  $\hat{\mathbf{R}}_{\text{new}} = \mathbf{Z}_{\text{new}}\mathbf{Z}_{\text{new}}^{\text{H}}$ is the empirical covariance of $\mathbf{Z}_{\text{new}}$. Intuitively, the GLRT solution scans the spatial (covariance) domain on blocks of time samples and results in a contrast peak (further shown and discussed in Sec. III) when a signal (i.e., the SOI) impinging the antenna array from a different direction than the interference turns on in $\hat{\mathbf{R}}_{\text{new}}$ that was not present in $\hat{\mathbf{R}}_{\text{old}}$.

\subsection{GLRT Breakdown in Nonlinearity}
Now, let us consider the effect of the GLRT test statistic given in (\ref{detection}) when a nonlinearity is induced on the received signal. In this case, we will assume that we receive $f(\mathbf{Z}): \mathbb{C}^{M \times L} \rightarrow \mathbb{C}^{M \times L}$ instead of $\mathbf{Z}$, where $f(\cdot)$ is some unknown nonlinear function that is individually applied to each element in $\mathbf{Z}$. Specifically, the nonlinearity effect is given by
\begin{equation}
    f(\mathbf{Z}) = f([\mathbf{z}(1), \ldots, \mathbf{z}(L)]) = [f(\mathbf{z}(1)), \ldots, f(\mathbf{z}(L))]. 
\end{equation}
The element-wise effect of $f(\cdot)$ is critical because it will affect each sample differently. For example, certain received time samples in $\mathbf{Z}$ will remain in the linear operating region and thus will not be significantly distorted by $f(\cdot)$, whereas other samples will experience high degrees of clipping and distortion due to the nonlinearity. The clipping distortion of each antenna element can be observed in the eigenvalue spread of the covariance matrix, $\hat{\mathbf{R}}_{\text{old}}$. 

The induced nonlinearity directly changes the distribution of $\mathbf{Z}$ and, as a result, the initial assumptions about the distributions of the interference and SOI in (\ref{hyp_def}) no longer hold. Furthermore, since the nonlinear function, $f(\cdot)$, is unknown, we are unable to characterize the distribution of the signal after it has been transformed via $f(\cdot)$, thus preventing us from deriving a new GLRT-based test statistic to operate on $f(\mathbf{Z})$. The inability to analytically characterize nonlinearly-processed received signals motivates a data driven approach to signal detection (as described in Sec. II-D), where data samples can be augmented such that the GLRT is able to operate accurately. 


\subsection{DNN-Based Nonlinear Sample Classification}

To combat the breakdown of the GLRT solution on nonlinear signals, we implement a dense (fully connected) neural network (DNN) classification model, which employs the time samples comprising $f(\mathbf{Z})$ for training. Specifically, our objective is to train the DNN to classify excessively nonlinear time samples, i.e., $f(\mathbf{z}(t))$, using both a linear received signal, $\mathbf{Z}$, and its corresponding nonlinear transformation, $f(\mathbf{Z})$. At test time, when we only receive $f(\mathbf{Z})$ and are unaware of $\mathbf{Z}$, we will use the DNN to identify and eliminate excessively nonlinear samples from $f(\mathbf{Z})$ before applying the GLRT test statistic for detection. Our overall DNN-prepended GLRT proposed framework is shown in Fig. \ref{block_diagram}.

For compatibility with real-valued DNNs, we will represent each time sample (column vector) as a real-valued vector consisting of the sample's corresponding real and imaginary components. Specifically, we denote $f(\mathbf{z}(t))$ as $f(\mathbf{z}(t)) \in \mathbb{R}^{2M \times 1}$ with the form
\begin{equation} \label{reshape}
f(\mathbf{z}(t)) = 
    \begin{bmatrix}
        \Re \{f(\mathbf{z}_{1}(t))\} \\
        \vdots  \\ 
        \Re \{f(\mathbf{z}_{M}(t))\}  \\
        \Im \{f(\mathbf{z}_{1}(t))\} \\
        \vdots  \\ 
        \Im \{f(\mathbf{z}_{M}(t))\} \\
    \end{bmatrix}, 
\end{equation} 
where $\Re \{\cdot\}$ and $\Im \{\cdot\}$ denote the real and imaginary components of $\{\cdot\}$ and the subscript denotes the antenna element from which the sample was collected at time $t$. 


Using a set of training data for which we have both the linear and nonlinear representation of the received signal, $\mathbf{z}(t)$  and $f(\mathbf{z}(t)) \hspace{1mm} \forall \hspace{1mm} t$, we measure the Euclidean distance between each time sample, i.e., between $\mathbf{z}(t)$ and $f(\mathbf{z}(t))$, and populate a one-hot label vector $\mathbf{y}(t) \in \mathbb{R}^{2 \times 1}$ for each sample. 
Specifically, for each time sample, $t$, we compute 
\begin{equation}
    d_{t} = ||\mathbf{z}(t) - f(\mathbf{z}(t))||_{2}, 
\end{equation}
with the corresponding label determined as
\begin{equation} \label{labels}
  \mathbf{y}(t) =
    \begin{cases}
      [1, 0]^{\text{T}}, & d_{t} \leq d_{\text{T}}\\
      [0, 1]^{\text{T}}, & d_{t} > d_{\text{T}}
    \end{cases},        
\end{equation}
where $d_{\text{T}}$ is an empirically determined threshold from the training data, and $\mathbf{y}(t)$ indicates whether the corresponding time sample is excessively nonlinear.

Next, we train a DNN, $\zeta: \mathbb{R}^{2M \times 1} \rightarrow \mathbb{R}^{2 \times 1}$, using $f(\mathbf{z}(t))$ and $\mathbf{y}(t) \hspace{1mm} \forall \hspace{1mm} t$ to classify the time samples, $f(\mathbf{z}(t))$, that are highly distorted from their linear counterparts. Note that the input to the DNN is $f(\mathbf{z}(t))$ since, during test time, we will only receive the nonlinear signal, and thus need to classify its time samples that have contributed to the nonlinearity to the greatest extent. We consider a $p$-layered DNN, consisting of $A$ units each, with a softmax output layer yielding the DNN prediction $[\hat{\mathbf{y}}_{1}(t), \hat{\mathbf{y}}_{2}(t)]$, where the input is predicted to be nonlinear if $\hat{\mathbf{y}}_{1}(t) < \hat{\mathbf{y}}_{2}(t)$. In place of a two-dimensional softmax output, the DNN could output a one-dimensional sigmoidal value. However, we empirically found that the former setup yields a more accurate nonlinear sample classifier. 

Using $f(\mathbf{z}(t))$ and $\mathbf{y}(t) \hspace{1mm} \forall \hspace{1mm} t$ as training samples, the DNN minimizes the mean squared error loss function. At test time, each column (i.e., time sample) is forward propagated through the DNN, and time samples classified as highly distorted due to the nonlinearity are deleted. The edited matrices of $f(\mathbf{Z}_{\text{old}})$ and $f(\mathbf{Z}_{\text{new}})$, denoted as $\mathbf{Z}_{\text{old}}^{\prime}$ and $\mathbf{Z}_{\text{new}}^{\prime}$, respectively, correspond to $f(\mathbf{Z}_{\text{old}})$ and $f(\mathbf{Z}_{\text{new}})$ after their time samples that were classified as nonlinear by $\zeta$ were removed. $\mathbf{Z}_{\text{old}}^{\prime}$ and $\mathbf{Z}_{\text{new}}^{\prime}$ are then used to compute the GLRT test statistic in (\ref{detection}).


\vspace{-0.25cm}

\subsection{Complexity Analysis}
\vspace{-0.1cm}

Here, we analyze the computational complexity of our proposed framework during inference. The GLRT statistic in (\ref{detection}) consists of $2$ $M \times k$ matrix Hermitian operations and $2$ $M \times k$ matrix multiplications followed by an $M \times M$ matrix inverse and an $M \times M$ matrix multiplication. Thus, the computational complexity of the GLRT is $\mathcal{O}(k^{2} +M^{2})$. Since the DNN pre-pended to the GLRT in our proposed framework consists of $p$ dense layers with $A$ units each, it adds an additional complexity of $\mathcal{O}(2(p+1)A^{3})$. This yields a total time complexity, for our method, of $\mathcal{O}(pA^{3} + k^{2} +M^{2})$. During inference, our method requires roughly $60$ ms to compute the test statistic in (\ref{detection}) for the simulation environment considered in Sec. III. Thus, despite the overhead incurred from the DNN, our method remains computationally feasible for practical use.


\section{Simulations}




\subsection{Signal Generation}

For our empirical evaluation, we generate an interference signal, $a_{1}$, and a SOI, $a_{2}$. Each signal contains a sequence of random bits, is modulated according to the BPSK constellation, and transmitted with a carrier frequency of $f_{c} = 10$ MHz and a bandwidth of $1$ MHz. At the receiver, we use $M = 4$ antenna elements and consider a received signal with an observation window of $L = 2000$ time samples. In addition, we use $k = 48$ time samples to construct each initial instance of $\mathbf{Z}_{\text{old}}$ and $\mathbf{Z}_{\text{new}}$. Here, we consider two setups by scaling the AWGN: (a) an interference to noise ratio (INR) of 40 dB and a signal to noise ratio (SNR) of 15 dB (corresponding to perhaps an RF application); and (b) an INR of 57 dB and an SNR of 37 dB (corresponding to perhaps an audio application). The large power difference between the SOI and interference introduces an additional challenge for detection, which is achievable in the linear case but much more challenging in nonlinearity. 

In our analysis, we use the hyperbolic tangent function as the induced nonlinearity, i.e., we define $f(\mathbf{Z}) = \text{tanh}(\mathbf{Z})$. This function is a representative model for hardware nonlinearities, since it consists of a linear and a nonlinear region, thus reflecting the effect of amplifiers when only certain received samples experience nonlinear saturation \cite{tanh_nl}. 


\vspace{-0.25cm}
\subsection{DNN Implementation}


The DNN is trained on a generated received signal $\mathbf{Z}_{\text{Train}}$ and its nonlinear counterpart $f(\mathbf{Z}_{\text{Train}})$ to identify excessively nonlinear signals based on the method in Sec. II-D. The employed DNN contains $p = 3$ layers with $A = 10$ hyperbolic tangent units each. During training, we use a batch size of 64, a learning rate of $0.001$, and 500 epochs with a patience on the training loss of 25 epochs as the stopping criterion. 


\vspace{-0.25cm}
\subsection{Detector Performance}


\begin{figure}[t]
\centering
{\includegraphics[scale=0.2]{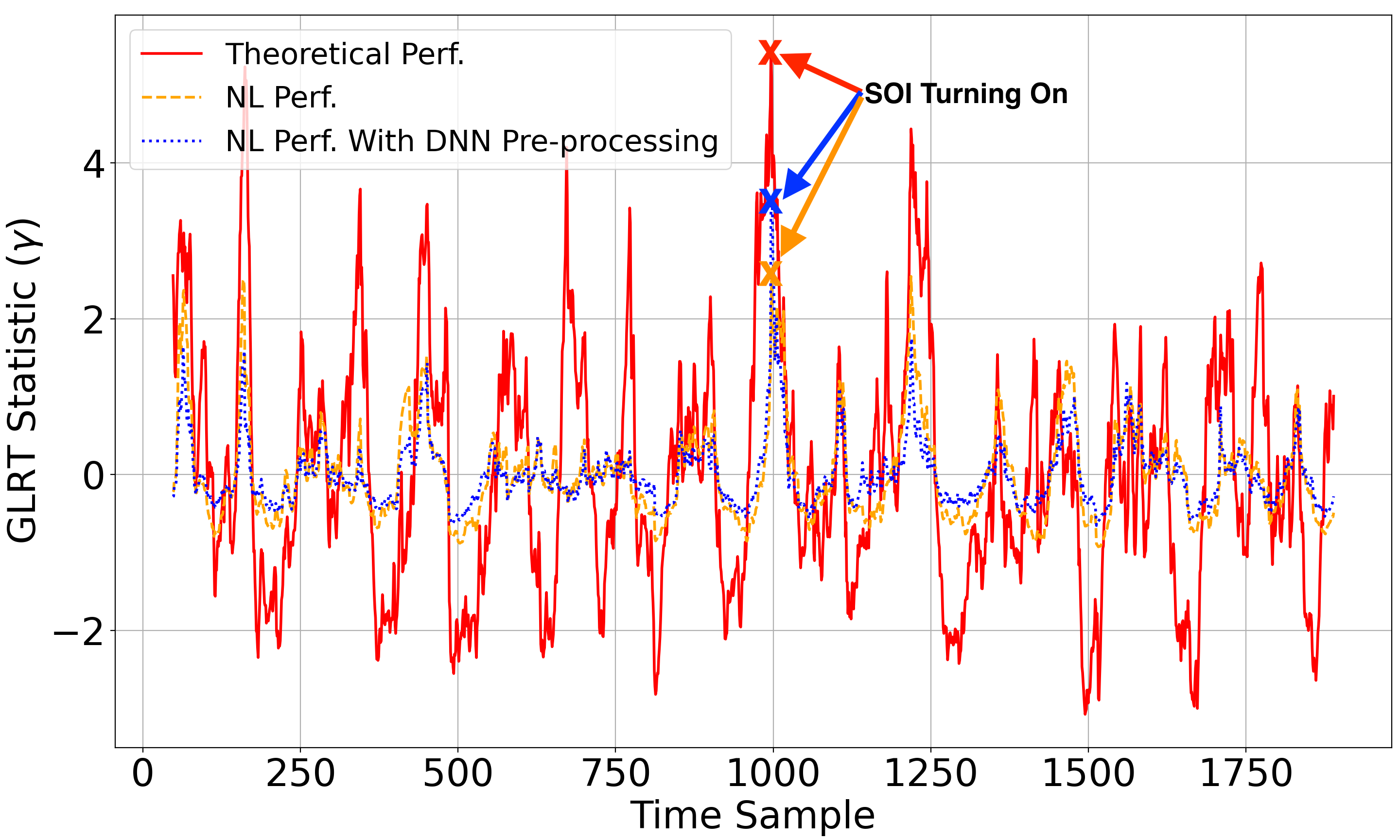}}
{\includegraphics[scale=0.2]{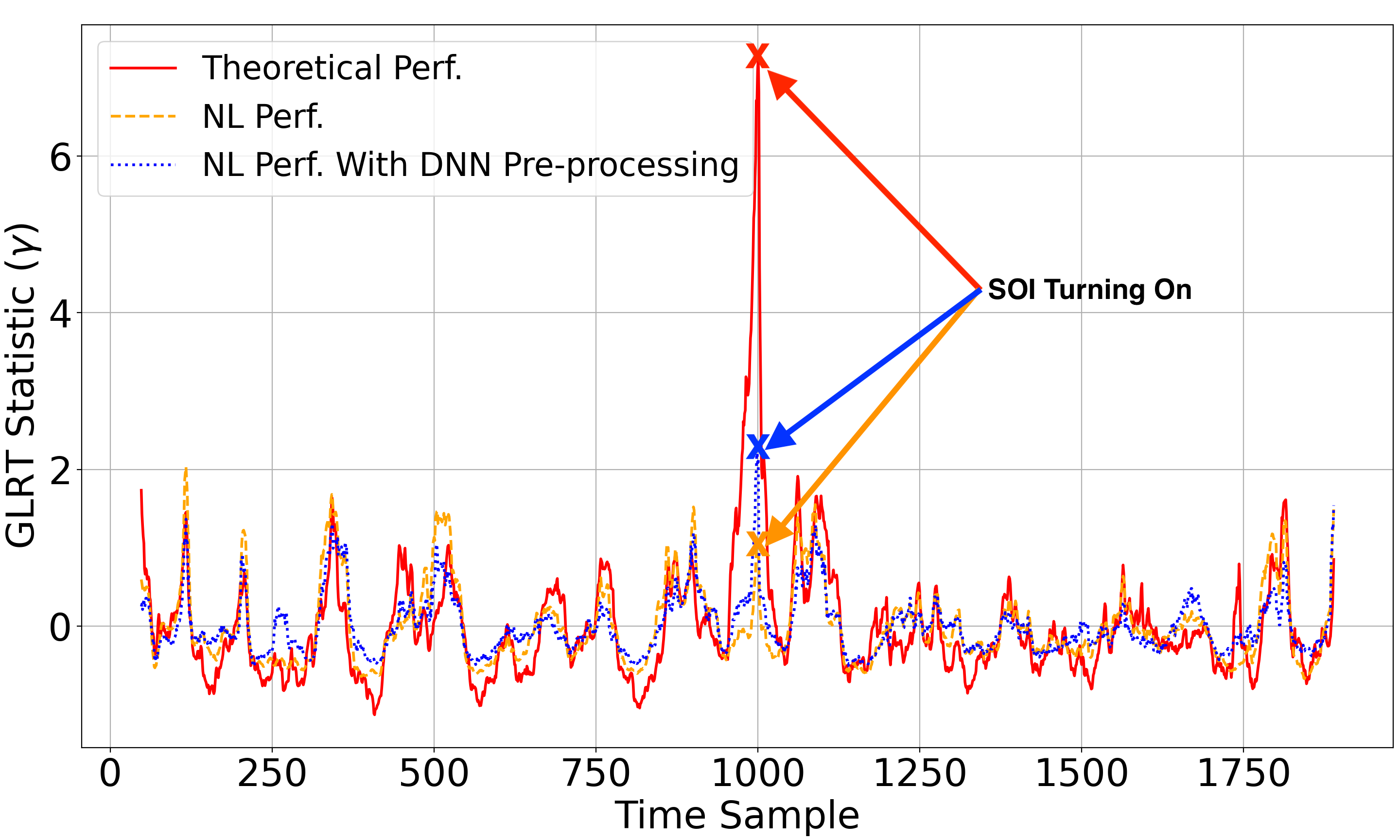}}
\caption{\small GLRT performance of a SOI turning on at $T_{0} = 1000$ in setups (a) (top) and (b) (bottom). The contrast peak, produced by the SOI turning on at $T_{0}$ in the ideal case, is not distinctive in the nonlinear case but reappears after using DNN pre-processing.}
\label{metric}
\end{figure}

\begin{figure}[t]
\centering
{\includegraphics[scale=0.2]{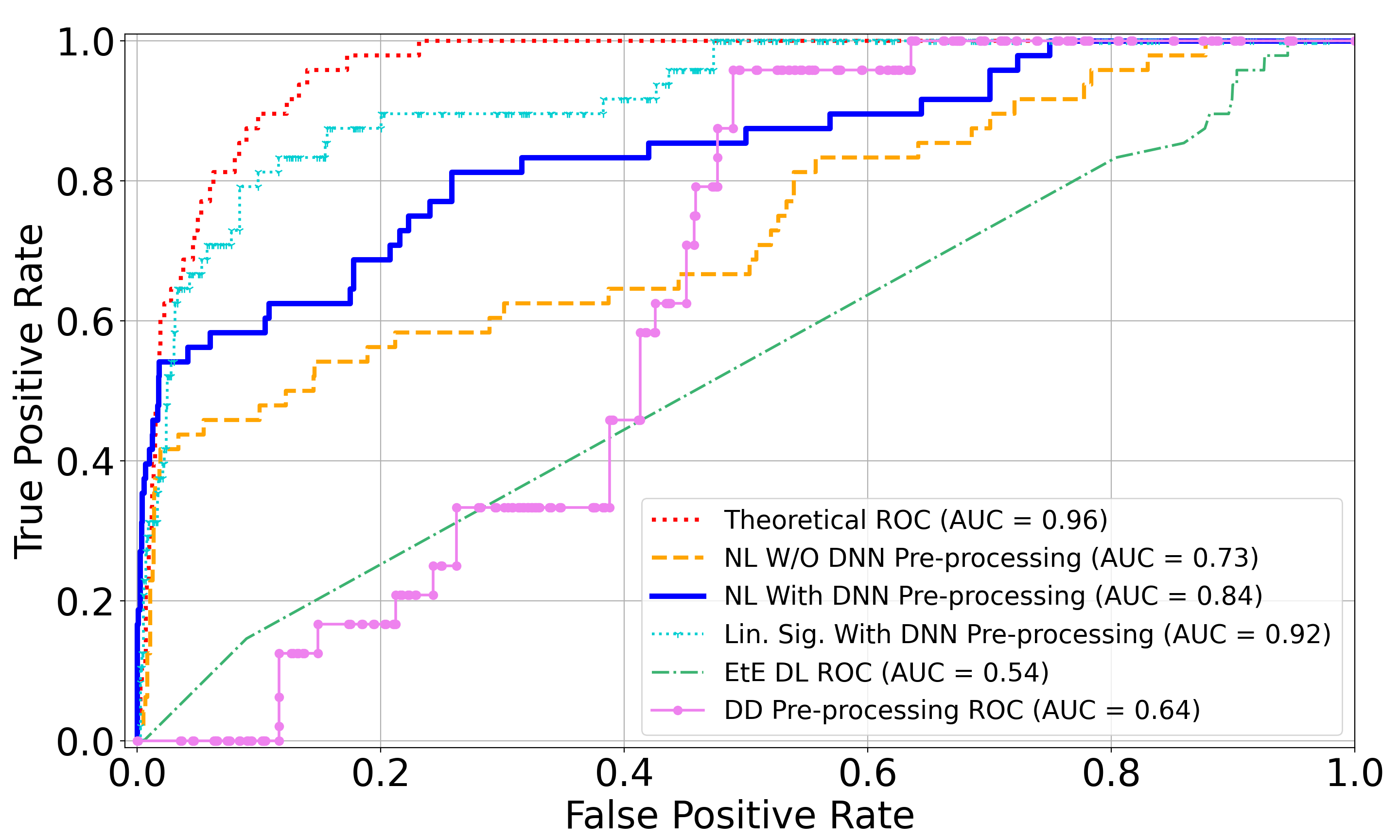}}
{\includegraphics[scale=0.2]{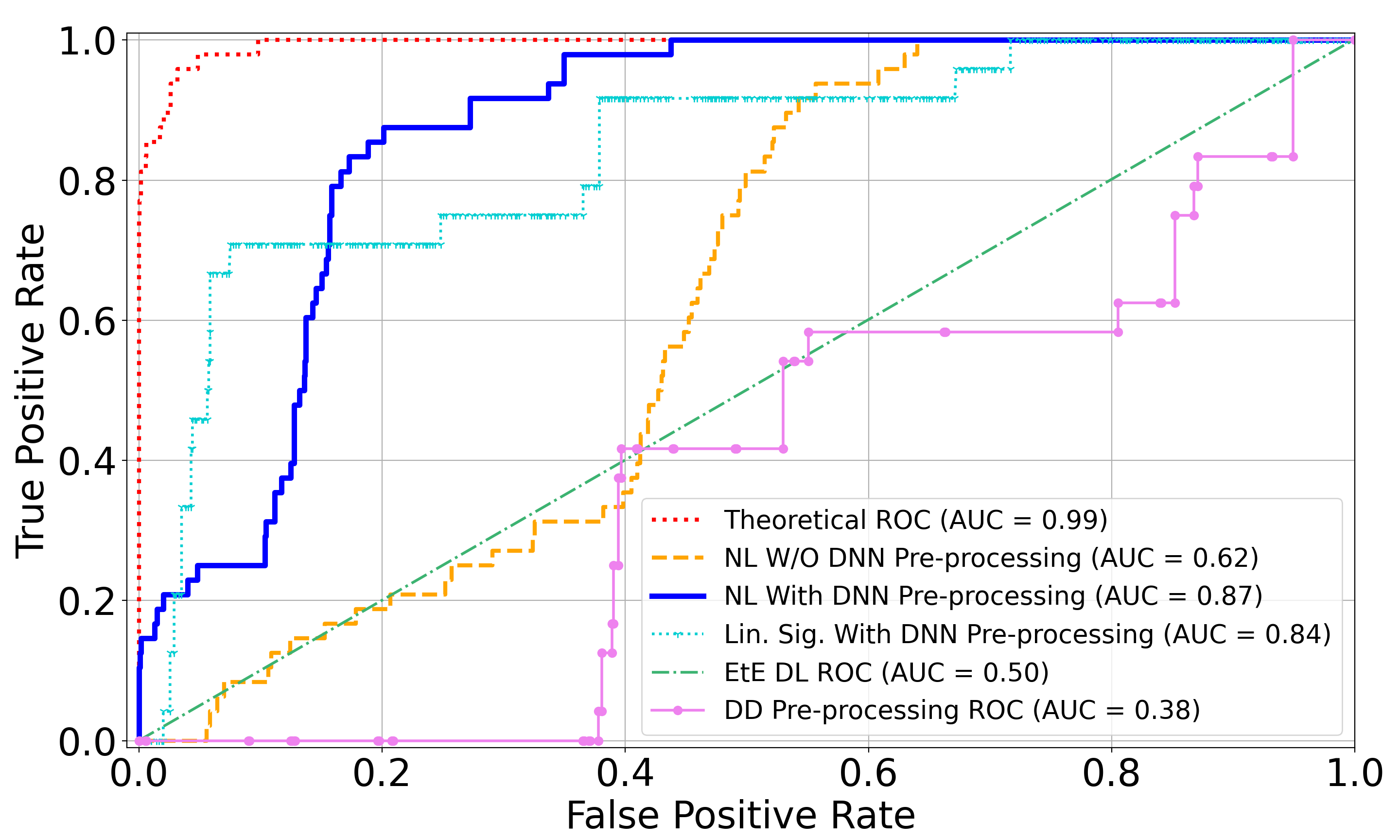}}
\caption{\small Detection performance in each setups (a) (top) and (b) (bottom) of the theoretical signal, NL signal, linear and NL signal with our proposed DNN pre-processing, and two baselines. Our methodology significantly improves the detection performance under nonlinearity in comparison to the GLRT operating solely on the NL signal and over the considered baselines.}
\label{roc}
\end{figure}

We evaluate our methodology on $f(\mathbf{Z}_{\text{Test}})$, and in addition, we generate the linear counterpart, $\mathbf{Z}_{\text{Test}}$, to compare the nonlinear and linear detection performance. During implementation, we scale the nonlinearity based on a factor $\alpha$ as $f(\mathbf{\alpha Z}) / \alpha$ to achieve a desirable nonlinearity in each considered setup.
Fig. \ref{metric} shows the GLRT test statistic on (i) a theoretical (linear) signal, (ii) a nonlinear signal, and (iii) the same nonlinear signal after DNN pre-processing for setups (a) and (b). The SOI turns on at $T_{0} = 1000$. As shown by the metric on the linear signal, a distinctive contrast peak is captured at the time in which the SOI turns on in the received transmission. When the signal undergoes a nonlinearity, and $f(\mathbf{Z}_{\text{Test}})$ is received without DNN pre-processing, several false alarm peaks are produced at multiple incorrect time samples, and no distinctive contrast peak is captured at $T_{0}$ when the SOI turns on. When our proposed methodology is applied to the nonlinear signal, however, the GLRT test statistic produces a distinctive contrast peak correctly at $T_{0} = L/2$, albeit to somewhat of a lesser degree than in the theoretical case. This behavior stems from the remaining samples in $\mathbf{Z}_{\text{old}}^{\prime}$ and $\mathbf{Z}_{\text{new}}^{\prime}$ still containing nonlinear samples. Using the curves in Fig. \ref{metric}, we vary the detection threshold and measure the resulting false positive versus true positive detection rates to obtain the performance curves shown next in Fig. \ref{roc}. 

Fig. \ref{roc} shows the receiver operating characteristic (ROC) detection performance curves of the GLRT for each setup. In addition, we compare our proposed method to two baselines: (i) end-to-end deep learning (EtE DL) \cite{dl2}, where a deep learning classifier is trained on both linear and nonlinear signals to perform detection directly in place of the GLRT; and (ii) data driven (DD) pre-processing \cite{dd_pp}, where the received nonlinear signal is first pre-processed in an attempt to recover the linear signal prior to applying the GLRT. In our adoption of \cite{dd_pp}, we train an autoencoder (AE) to map nonlinear signals (taken as input to the AE) to their linear counterparts (given as output by the AE). The GLRT test statistic in (\ref{detection}) is then applied on the AE output. Finally, for completion, we also employ our proposed DNN pre-processing methodology on a linear signal to show its effect when it is applied to a signal that does not suffer from nonlinear distortions.

From Fig. \ref{roc}, we see that our method is able to improve the area under the curve (AUC) of the ROC curve in both cases over using the nonlinear signal without DNN pre-processing and the other considered baselines. The lower AUC, in comparison to the theoretical linear case, is attributed to the increased magnitudes of the false positive contrast peaks produced by the GLRT in Fig. \ref{metric}. In particular, we note that our method significantly improves the true positive detection rate at low false alarm levels, which is often desirable in the context of wireless signal detection. The nearly random performance of the EtE DL classifier is attributed to the large power difference between the interference and SOI, resulting in a lack of salient features for the classifier to perform detection on. Similarly, DD pre-processing faces challenges in the training step since the same nonlinear clipped time sample corresponds to several linear time samples (i.e., a one-to-many mapping), thus resulting in poorly trained AE models. Our proposed method avoids the sensitivity of mapping non-linear time samples to an approximation of their linear counterparts by instead classifying and removing nonlinear samples that degrade detection performance and then applying the GLRT. Finally, we see that applying our proposed framework to a linear signal only slightly degrades performance in comparison to the theoretical linear performance. However, since we can deduce when a received signal has undergone non-linear distortions from hardware constraints, we can employ our proposed framework on non-linear signals while resorting to the classical GLRT on linear observations.



\section{Conclusion and Future Work}

In this work, we proposed prepending a GLRT detector with a DNN in order to perform reliable detection in the presence of nonlinear distortions. We showed that our method was capable of performing accurate detection despite the nonlinearity, whereas nonlinear signals without any pre-processing were unable to be used for detection using the GLRT. In future work, we anticipate exploring additional nonlinear environments  that our framework could be applied to, including the extended target case \cite{ext_targets}. In addition, we expect to expand this work to dynamic environments where the direction of the interference or SOI may not be stationary and require the utilization of the GLRT in an FDA-MIMO \cite{fda_mimo} framework.



\bibliography{references}

\bibliographystyle{IEEEtran}

\end{document}